\documentclass[aps,prl,amsmath,amssymb,floatfix,twocolumn, amsmath, superscriptaddress, twocolumn]{revtex4}
\usepackage{multirow}
\usepackage{bbold}
\usepackage{subfigure}
\usepackage{color}
\usepackage{mathrsfs}
\usepackage{hyperref}
\usepackage[normalem]{ulem}

\usepackage{amsfonts, relsize, color}
\usepackage{graphics}
\usepackage{graphicx}
\usepackage{subfigure}
\usepackage{hyperref}
\usepackage{color}

\begin{document}
\title{\bf Interacting Weyl fermions: Phases, phase transitions, and global phase diagram}

\author{Bitan Roy}
\affiliation{Condensed Matter Theory Center and Joint Quantum Institute, Department of Physics, University of Maryland, College Park, Maryland 20742- 4111, USA}
\affiliation{Department of Physics and Astronomy, Rice University, Houston, Texas 77005, USA}

\author{Pallab Goswami}
\affiliation{Condensed Matter Theory Center and Joint Quantum Institute, Department of Physics, University of Maryland, College Park, Maryland 20742- 4111 USA}

\author{Vladimir Juri\v ci\' c}
\affiliation{Nordita,  Center for Quantum Materials,  KTH Royal Institute of Technology and Stockholm University, Roslagstullsbacken 23,  10691 Stockholm,  Sweden}

\date{\today}
\begin{abstract}
We study the effects of short-range interactions on a generalized three-dimensional Weyl semimetal, where the band touching points act as the (anti)monopoles of Abelian Berry curvature of strength $n$. We show that any local interaction has a \emph{negative} scaling dimension $-2/n$.  Consequently all Weyl semimetals are stable against weak short-range interactions. For sufficiently strong interactions, we demonstrate that the Weyl semimetal either undergoes a first order transition into a band insulator or a continuous transition into a symmetry breaking phase. A translational symmetry breaking axion insulator and a rotational symmetry breaking semimetal are two prominent candidates for the broken symmetry phase. At the one-loop order, the correlation length exponent for continuous transitions is $\nu=n/2$, indicating their non-Gaussian nature for any $n>1$. We also discuss the scaling of the thermodynamic and transport quantities in general Weyl semimetals as well as inside the broken symmetry phases.
\end{abstract}

\maketitle

\emph{Introduction}: There is tremendous ongoing interest in three dimensional, gapless topological systems~\cite{balatsky, RyuTeo, tanmoy-RMP, newfermions, kane-prb, volovik-book}. An interesting class of such systems is described by the isolated nodal points of time reversal or inversion symmetry breaking materials, where two non-degenerate bands touch~\cite{vishwanath, balents, Fang-HgCrSe, bergevig, nagaosa, hasan-DWSM, cava-Todd, choiresku-Todd, balent-DWSM, volovik, goswami-balicas, roy-goswami-dssarma, sigrist, cdas, nabi, taas-2, nbas-1, yazdani, zunger}. These points are known as Weyl points, which act as the (anti)monopoles or (anti)hedgehog of Abelian Berry curvature. For a general monopole strength $n$ (with only $n=1,2,3$ in crystalline systems), the dispersion relations around the nodal points acquire the form $\epsilon_\pm(\mathbf{k}) \sim \pm \sqrt{v^2 k^2_z + \alpha^2_n k^{2n}_{\perp}}$, which can be observed in angle-resolved photoemission spectroscopy (ARPES) experiments~\cite{taas-2, nbas-1}, where $k^2_\perp=k^2_x+k^2_y$, and $\pm$ respectively denote the conduction and valence bands~\cite{Fang-HgCrSe, bergevig, nagaosa}. Notice that only the conventional Weyl point with $n=1$ features linear dispersion along all three spatial directions. A general Weyl semimetal (WSM) gives rise to $n$ Fermi arcs as the topologically protected zero energy surface states that can be seen in Fourier-transformed scanning tunneling microscopy (STM)~\cite{yazdani}. Currently, there are proposals for double WSMs in HgCr$_2$Se$_4$~\cite{Fang-HgCrSe, bergevig}, SrSi$_2$~\cite{hasan-DWSM}, the ferromagnetic phase of pyrochlore iridates~\cite{balent-DWSM}, and time reversal symmetry breaking, chiral superconducting states of $^3$He-A~\cite{volovik}, URu$_2$Si$_2$, UPt$_3$~\cite{goswami-balicas} and SrPtAs~\cite{sigrist}, and triple WSMs in molybdenum mono-chalcogenide compounds A(MoX)$_3$ (A=Rb, Tl; X=Te)~\cite{zunger}. There is also growing interest in understanding the effects of interactions on topological semimetals~\cite{aji, zhang-CDW, nandkishore, nomura, roy-catalysis, sukhachov, buividovich, dassarma, Wang-Pe, hughes, thomale, yao-nematic, goswami-chakravarty, hosur, nagaosa-isobe, gonzalez, throckmorton, hsin-hua, hongyao, roy-JHEP, katsnelson, y-ran}. However, the effects of generic short-range interactions on the global phase diagram of a general WSM are not well understood yet and we investigate this problem in the current Rapid Communication.

\begin{figure}
\includegraphics[width=4.2cm, height=4.2cm]{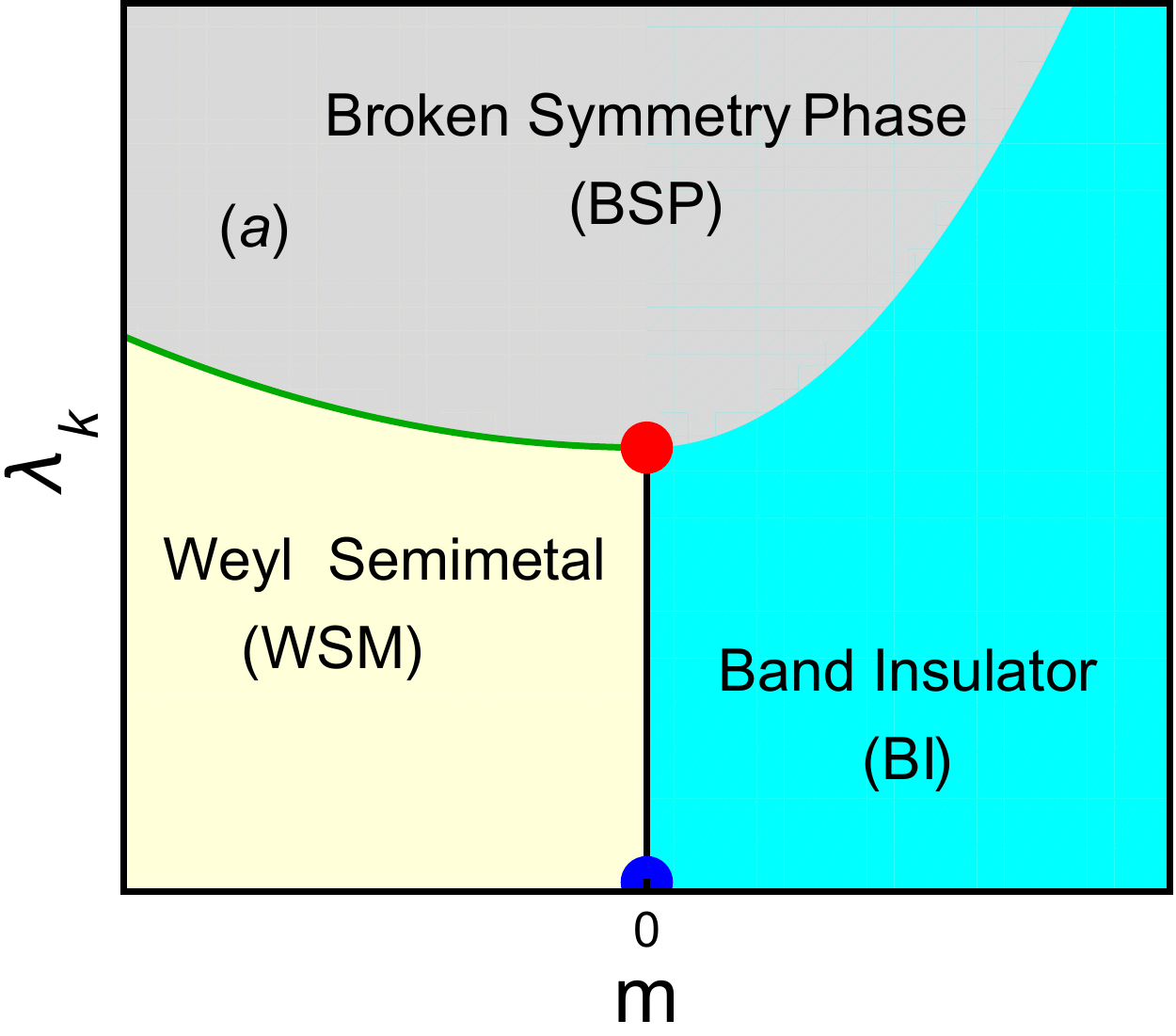}
\includegraphics[width=4.2cm, height=4.25cm]{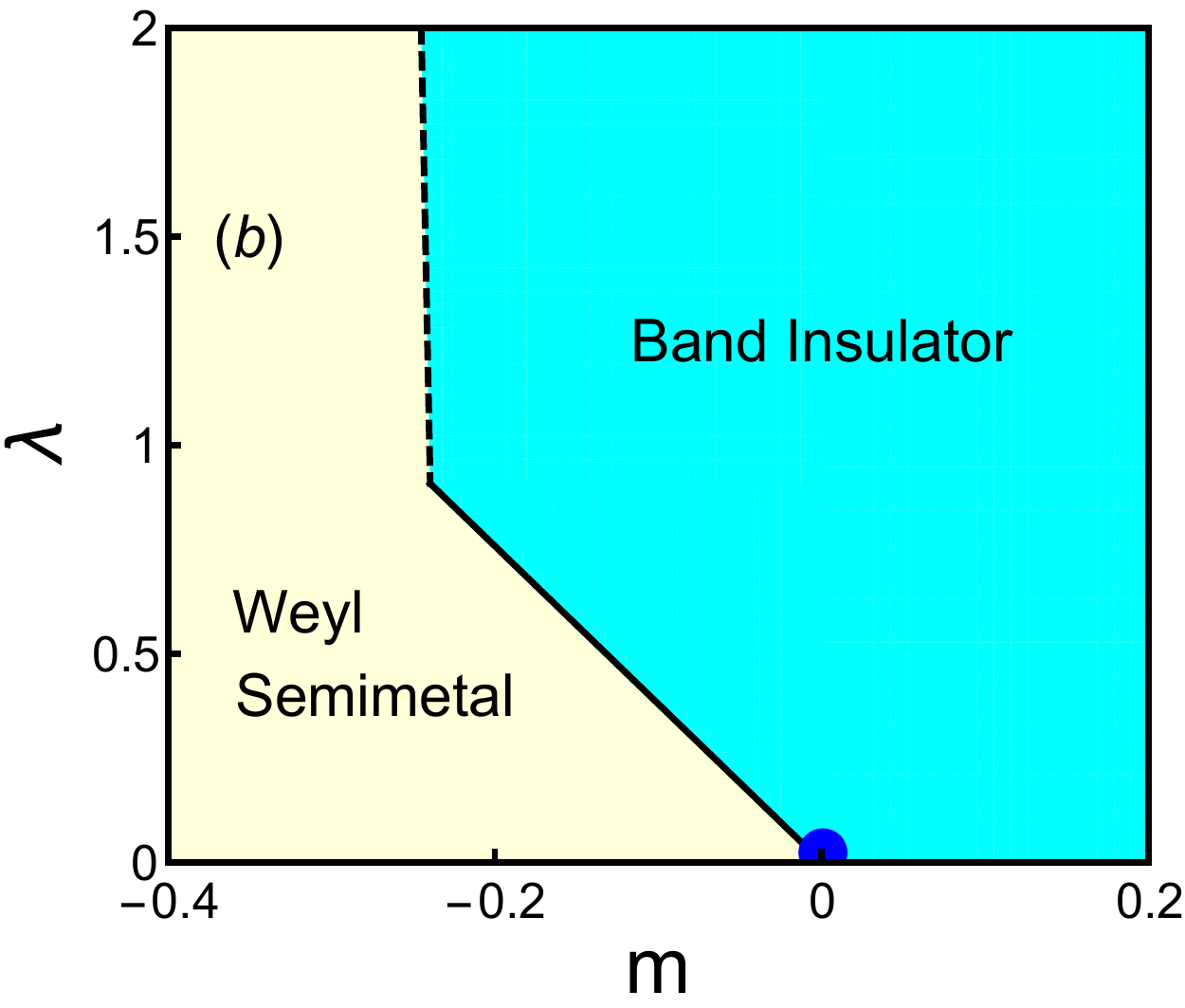}
\caption[]{Representative phase diagrams of an interacting Weyl system. Here $\lambda_k$s (with $k=2,3,4$) are dimensionless interaction couplings for Weyl fermions [see Eqs.~(\ref{int_Hamil}) and (\ref{RG_epsilon})] and $\lambda$ is that for critical excitations, residing at the WSM-BI QCP (blue dot), located at $m(=\Delta/E_\Lambda)=0$, the dimensionless band parameter, where $E_\Lambda$ is the ultraviolet energy cutoff of critical excitations. For weak interactions, the transition between WSM and BI across the black solid lines is always continuous. By contrast, sufficiently strong interactions can either (a) eliminate the direct WSM-BI transition through nucleation of a BSP (see the lower panel of Fig.~\ref{competing_PD} for possible BSPs) or (b) cause a direct first order transition (across the dashed line) between the WSM and BI [obtained from Eq.~(\ref{freeenergy}) for $n=1$]. We emphasize that no symmetry is spontaneously broken across the direct WSM-BI transitions. The BI-BSP as well as WSM-BSP transitions are continuous. The red dot represents a \emph{multicritical point}, where WSM, BI and BSP meet. The direct first order transition between WSM and BI, shown in panel (b), can also be eliminated by a BSP phase. Here We do not consider a possible Mott localization of Weyl fermions for very large interactions. Panel (a) is schematic, and panel (b) for $n=2, 3$ are similar to the one shown above.  }\label{firstorder}
\end{figure}

At the generalized Weyl point, the density of states (DOS) vanishes as $\varrho(E) \sim |E|^{2/n}$, leading to a \emph{negative} scaling dimension $-2/n$ for all local four-fermion interactions. Consequently, we find that all WSMs are stable against infinitesimally weak, short-range interactions. Notice that local interactions are marginal perturbations for $n \to \infty$, as this hypothetical limit corresponds to linearly dispersing one-dimensional chiral fermions with constant DOS. The perturbative renormalization group (RG) calculations can thus be controlled by the parameter $\epsilon=2/n$ (about $n \to \infty$), following the spirit of $\epsilon$-expansion~\cite{zinn-justin}. From the RG analysis, we establish that a WSM with $n \geq 1$ can undergo a continuous quantum phase transition (QPT) into either a translational symmetry breaking axion insulator (AI) or a rotational symmetry breaking, gapless nematic phase. To one loop order, the correlation length exponent (CLE) for continuous QPTs is given by $\nu=n/2$. Therefore, strongly interacting general WSMs with $n>1$ can support rare examples of \emph{non-Gaussian itinerant quantum criticality in three dimensions}. We also find a limited parameter space, where sufficiently strong interactions can drive a \emph{first-order} QPT between a band insulator (BI) and the WSM, and there is no symmetry distinction between these two phases. Our main findings regarding the global phase diagram of an interacting WSM are illustrated by two types of phase diagrams displayed in Fig.~\ref{firstorder}. We also elucidate the nature of low energy excitations inside the broken symmetry phases (BSPs), and their imprints on experimentally measurable quantities. Altogether here we develop a unified field theoretic description (such as the $\epsilon$-expansion in terms of monopole charge $n$) of interacting general WSMs that manifests an intriguing confluence of nodal topology of Weyl fermions, exotic broken symmetry phases (such as AI and nematic orders), emergent quantum critical phenomena across continuous QPTs, and fluctuation-driven first-order transition.

\emph{Model}: All kinds of Weyl excitations can be obtained from an appropriate two-band model $H=\sum_{\mathbf k} \psi^\dagger_{\mathbf k} \; \left[ {\mathbf N} (\mathbf k) \cdot {\boldsymbol \sigma} \right]\; \psi_{\mathbf k}$, after suitably choosing the spin/pseudospin vector ${\mathbf N} (\mathbf k)$, where $\psi^\top_{\mathbf k}=\left( c_{{\mathbf k}, \uparrow}, c_{{\mathbf k}, \downarrow}\right)$ is a two-component spinor, $c_{{\mathbf k}, \alpha}$ is the fermion annihilation operator with momentum $\mathbf k$ and (pseudo)spin projection $\alpha=\uparrow, \downarrow$, and $\boldsymbol \sigma$ are Pauli matrices~\cite{roy-bera-sau, supplementary}. For our discussion of the low energy physics, it is sufficient to consider the following continuum model of a general WSM
\allowdisplaybreaks[4]
\begin{equation}\label{SM_QCP}
H_n=\alpha_n k^n_\perp \left[ \sigma_1 \cos(n\phi_{\mathbf k}) + \sigma_2 \sin(n\phi_{\mathbf k}) \right] + \sigma_3(B k^2_z+\Delta),
\end{equation}
where $\phi_{\mathbf k}=\tan^{-1}(k_y/k_x)$, and we have set $\hbar=1$ and lattice spacing $a=1$. For example, $\alpha_1$ bears the dimension of Fermi velocity, while $\alpha_2$ and $B$ have the dimensions of mass, and in this notation $N_1({\mathbf k})=\alpha_n k^n_\perp \cos(n\phi_{\mathbf k})$, $N_2({\mathbf k})=\alpha_n k^n_\perp \sin(n\phi_{\mathbf k})$, $N_3({\mathbf k})=B k^2_z+\Delta$. The sign of $\Delta$ does not change any underlying symmetry, but it gives rise to two distinct physical states. While $\Delta<0$ corresponds to the WSM phase, $\Delta>0$ leads to a BI (either Chern or trivial). The QCP between these two phases is located at $\Delta=0$.

The low-energy physics of Weyl fermions is captured by a \emph{four-component} spinor $\Psi^\top_{\mathbf k}=\big( c_{{\mathbf K}+ {\mathbf k},\uparrow},c_{{\mathbf K}+ {\mathbf k}, \downarrow},$ $c_{-{\mathbf K}+{\mathbf k}, \uparrow}, c_{-{\mathbf K}+{\mathbf k}, \downarrow} \big)$, after linearizing the theory around right ($+$) and left ($-$) handed Weyl nodes, respectively located at $\pm {\mathbf K}$, with ${\mathbf K}=(0,0,\sqrt{-\Delta/B})$. We find the low energy Hamiltonian for a general WSM with a monopole charge $n$ to be~\cite{supplementary}
\allowdisplaybreaks[4]
\begin{equation}
H^n_W=\alpha_n k^n_\perp \left[ \Gamma_{01} \cos(n\phi_{\mathbf k}) + \Gamma_{02} \sin(n\phi_{\mathbf k}) \right] +\Gamma_{03} v_z k_z,
\end{equation}
where $v_z=2\sqrt{- B \Delta}$, $\Gamma_{jk}=i \gamma_j \gamma_k$ and $\gamma_0=\tau_1 \otimes \sigma_3$, $\gamma_1=\tau_1 \otimes \sigma_2$, $\gamma_2=\tau_1 \otimes \sigma_1$, $\gamma_3=\tau_2 \otimes \sigma_0$. The Pauli matrices $\tau_\mu$ operate on the chirality index $\pm$. Notice that $[H^n_W,\gamma_5]=0$, where $\gamma_5 =\tau_3 \otimes \sigma_0$. Thus, $H^n_W$ possesses an \emph{emergent U(1) chiral symmetry} generated by $\gamma_5$, which  captures the \emph{translational symmetry} of the decoupled Weyl fermions~\cite{translation}. Now we analyze the form of effective short-range interactions by considering two types of gapless excitations, present inside WSM and at the QCP separating WSM and BI (see Fig.~\ref{firstorder}).

\emph{Interacting WSM}: In a WSM generic short-range interactions can be described by 16 interaction terms $H^W_{int}=g_{\mu \nu} \int d^3x \left( \Psi^{\dagger} \tau_\mu \otimes \sigma_\nu \Psi \right)^2$, where $\mu, \nu=0,1,2,3$. Eight of these couplings describe intranode or forward scattering, while the other eight correspond to internode or back scattering processes. After accounting for emergent chiral symmetry and rotational symmetry in the $xy$ plane (generated by $\Gamma_{12}$), and invoking the Fierz constraint~\cite{roy-fierz, supplementary}, we find that the effects of generic short-range interactions on a WSM can be addressed with only four independent coupling constants. We choose the interacting Hamiltonian as
\allowdisplaybreaks[4]
\begin{align}~\label{int_Hamil}
H_{int} &= \int d^3x \bigg\{ g_0 \left( \Psi^\dagger \Psi \right)^2 + g_4 \left[ \left( \Psi^\dagger \gamma_0 \Psi \right)^2 +\left( \Psi^\dagger \Gamma_{05} \Psi \right)^2 \right] \nonumber \\
&+ \sum^{2}_{j=1} \left[ g_2 \left( \Psi^\dagger \Gamma_{0j} \Psi \right)^2 + g_3 \left( \Psi^\dagger \Gamma_{3j} \Psi \right)^2 \right] \bigg\}.
\end{align}
Here, $g_0, g_2$ and $g_3$ correspond to forward scattering, while $g_4$ represents backscattering. The scaling dimension of quartic interactions is $[g_j]=-\frac{2}{n}$ (following the scaling of DOS). Thus, sufficiently weak short-range interaction is an irrelevant perturbation for any $n$.

\emph{Broken symmetry phases}: On the other hand, a general WSM can become unstable toward the formation of $(i)$ a translational or chiral symmetry breaking AI or $(ii)$ a rotational symmetry breaking \emph{nematic phase} at strong coupling [gray region in Fig.~\ref{firstorder}(a)], thereby eliminating the direct WSM-BI transition. As shown below the corresponding WSM-BSP QPTs turn out to be continuous.

 The AI order parameter hybridizes Weyl nodes of opposite chiralities (signifying a breakdown of translational symmetry), giving rise to a complex Dirac mass $H_{AI}= \Delta_{AI} \; \left[ \gamma_0 \cos \theta + \Gamma_{05} \sin \theta \right]$~\cite{y-ran, aji, zhang-CDW, nandkishore, nomura, roy-catalysis, sukhachov, buividovich, dassarma, Wang-Pe, hughes}. Within the mean-field picture, the uniformly gapped quasi-particle spectra become $E=\pm [\alpha^2_n k^{2n}_\perp+v^2 k^2_z +|\Delta_{AI}|^2]^{1/2}$~\cite{spectrum}. The U(1) Goldstone mode describing the fluctuations of $\theta$ is known as the \emph{axion} field, and it couples to the external electromagnetic field as $\theta \; \mathbf{E}\cdot \mathbf{B}$.

By contrast, a nematic order parameter does not mix two nodes and preserves the chiral symmetry. Due to the rotational symmetry breaking in the $xy$ plane, any nematic order splits a Weyl node of strength $n$ into $n$ copies of simple Weyl nodes with \emph{unit} monopole charge. However, the order parameter can couple to the right and left handed Weyl fermions with opposite or the same signs, respectively giving rise to $(a)$ an axial nematic (AN) or $(b)$ a regular nematic (RN) phase. The respective couplings between Weyl fermions and AN/RN order parameter fields are given by
\begin{equation}~\label{Nem-singleparticle}
H_{AN/RN}= \Delta_{AN/RN} \; \left[ \Gamma_{13/01} \cos \theta + \Gamma_{23/02} \sin \theta \right]. \nonumber
\end{equation}
Inside the AN phase, the emergent simple left ($L$) and right ($R$) handed Weyl nodes are respectively located at $k_\perp (\Delta_{AN})= \left[ \alpha^{-1}_n \Delta_{AN} \right]^{1/n}$ and $\phi_{\boldsymbol k}=\phi^{j}_{\boldsymbol k} \equiv \frac{\theta+ m_j \pi}{n}$, where $j=R, L$, $m_L=1,3, \cdots, 2n-1$ and $m_R=0,2, \cdots, 2n-2$. In contrast, the splitting of Weyl nodes near the left and right chiral points is identical inside the RN phase, and is given by $k_\perp (\Delta_{RN})$ and $\phi^{L}_{\boldsymbol k}$ or $\phi^{R}_{\boldsymbol k}$, depending on the sign of $\Delta_{RN}$~\cite{footnote-1}. Notice that both insulating and nematic orders suppress low energy DOS. While AI produces a hard gap in the quasi-particle spectrum, nematic orders cause a \emph{power-law} suppression of DOS $\varrho(E) \sim |E|^{\frac{2}{n}+1} \to |E|^2$ for $n>1$, whereas they only renormalize the location of the Weyl nodes, without altering the power-law dependence of DOS for $n=1$~\cite{roy-classification-BLG}. It is worth pointing out that the internal angle $\theta$ in AI, AN or RN phases in general gets locked by the crystal symmetry into a preferred set of values, which can only be found after incorporating higher gradient terms in the continuum Hamiltonian $H^n_W$. Next, we demonstrate the competition among these orders using RG analysis.

\begin{figure}
\includegraphics[width=7cm, height=4.2cm]{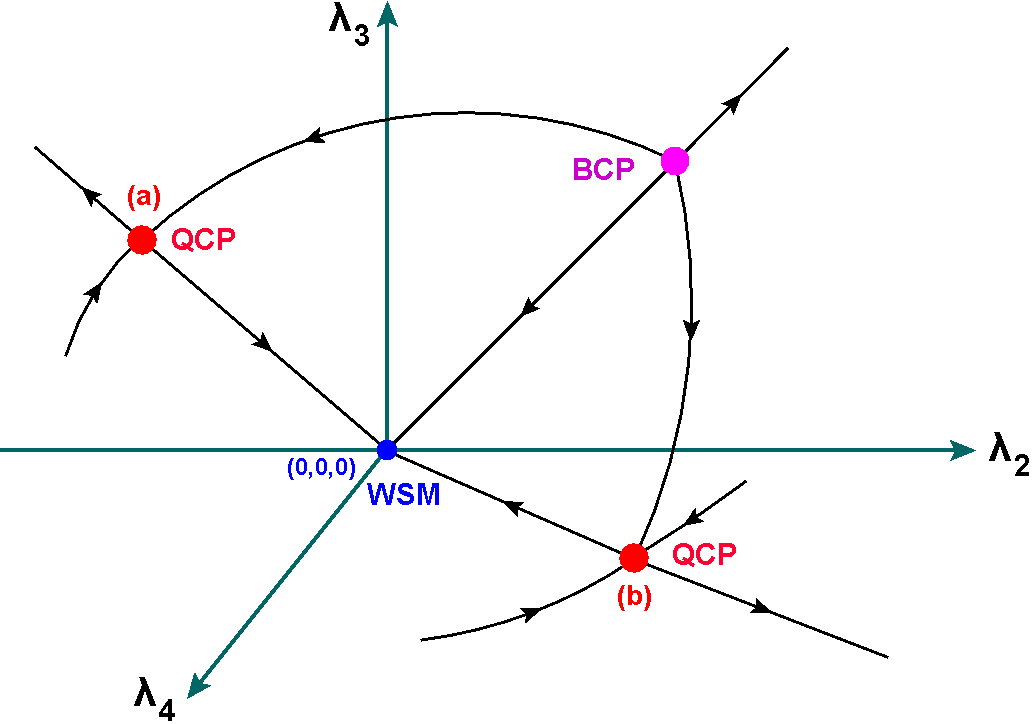}
\includegraphics[width=4.2cm, height=4.2cm]{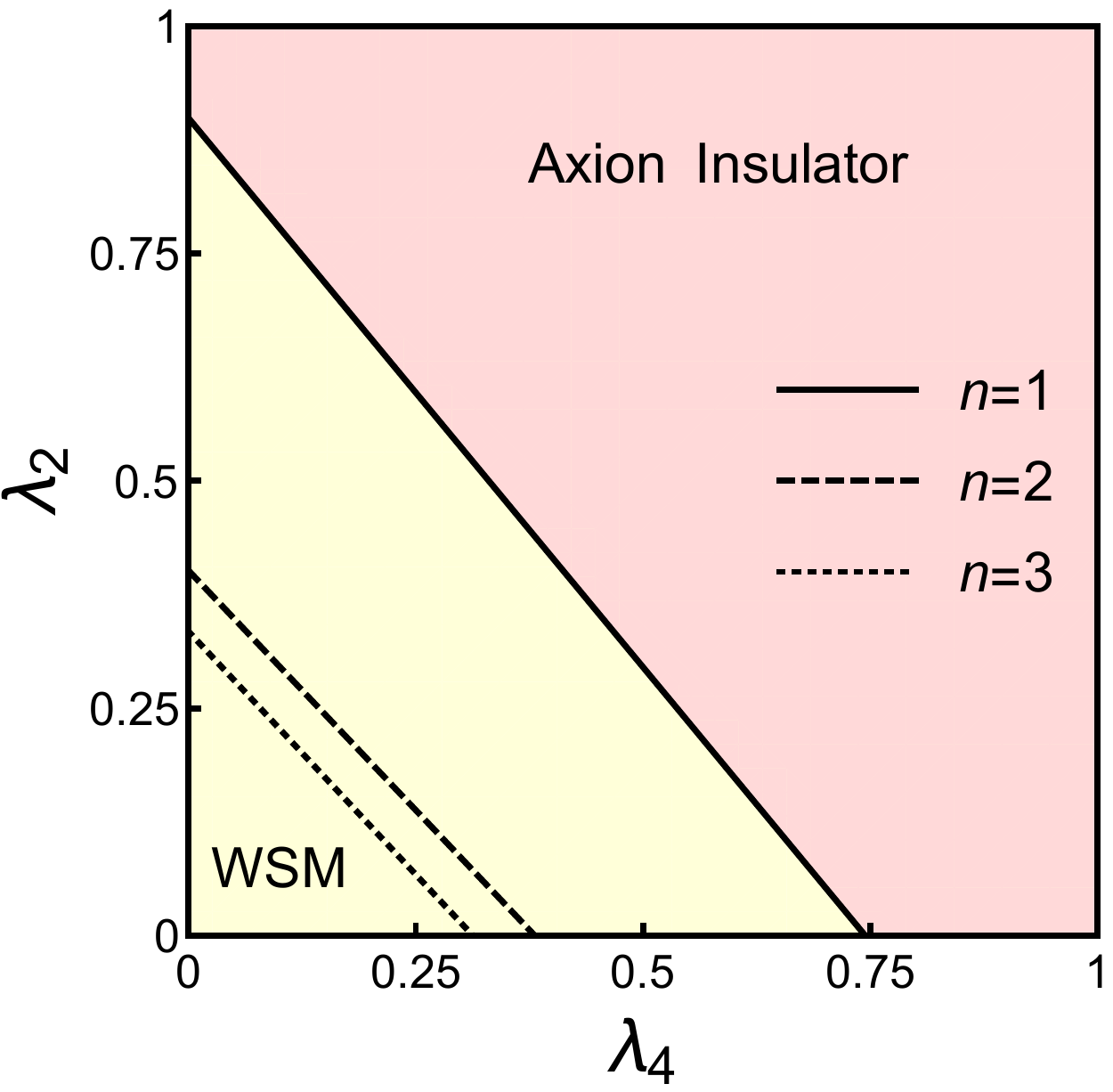}
\includegraphics[width=4.2cm, height=4.2cm]{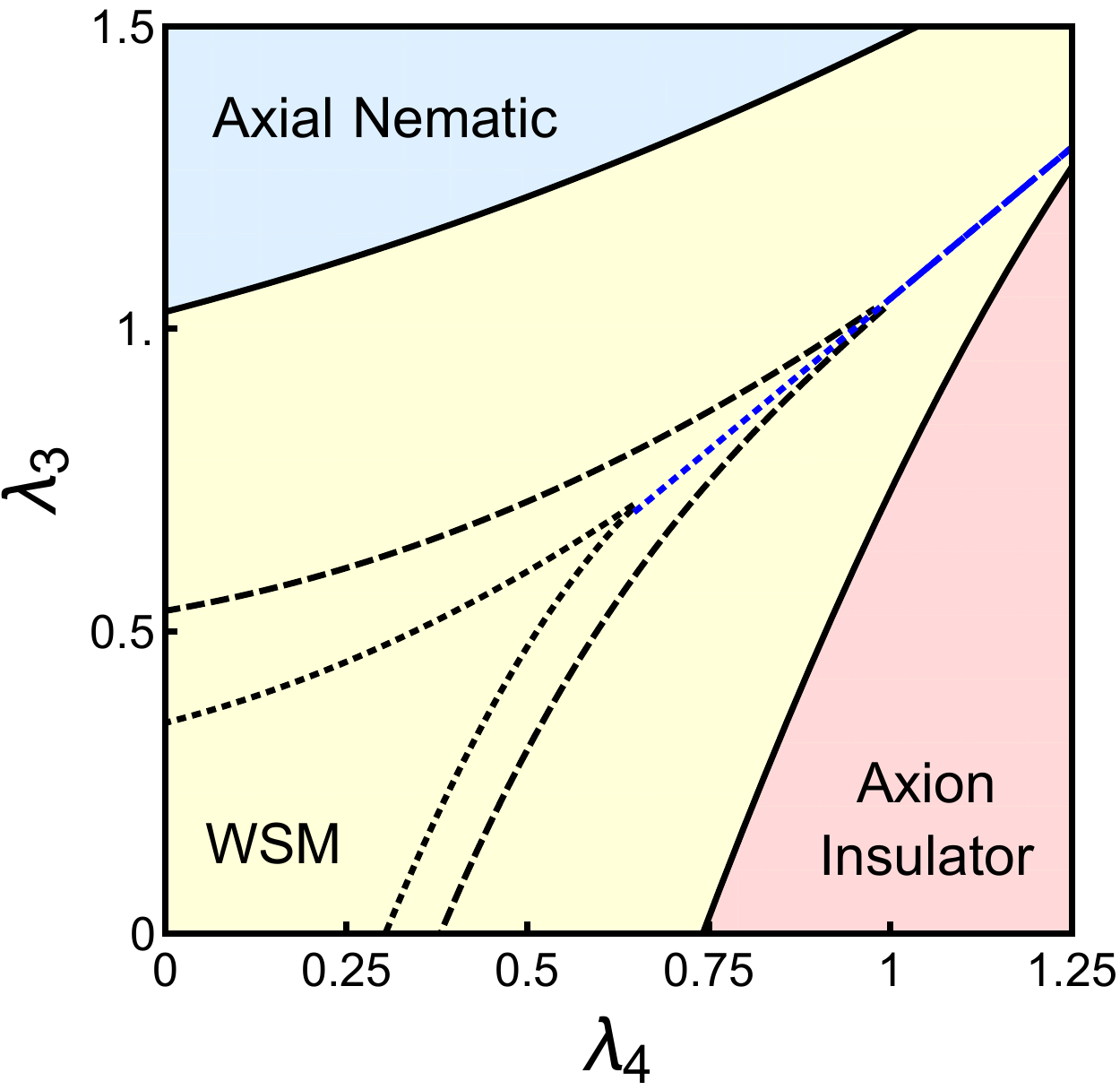}
\caption[]{Top: Schematic RG flow diagram in the $\left( \lambda_2, \lambda_3 ,\lambda_4 \right)$ space. Here, blue, magenta and red dots respectively represents fully stable and unstable bi-critical points (BCPs), and QCPs. Bottom: Representative cuts of phase diagram, depicting the instability of WSM toward broken symmetry phases [gray region in Fig.~\ref{firstorder}(a)] through continuous quantum phase transitions (across solid green lines) at strong coupling, as obtained by solving the RG flow equations~(\ref{RG_epsilon}) for $n=1$. The phase boundaries for $n=2 (3)$ are shown by dashed (dotted) lines. Across the blue line, the system undergoes a first-order transition between two distinct broken symmetry phases. Thus, with an increasing density of states the onset of broken symmetry phases occurs at weaker interactions. The phase diagram in $\left( \lambda_2, \lambda_3\right)$ plane is similar to the one shown in $\left( \lambda_4, \lambda_3\right)$ plane. At very strong interactions there can also be a Mott localization of Weyl fermions which is not considered here.
}\label{competing_PD}
\end{figure}

\emph{RG analysis}: Recall that local interactions are \emph{marginal} perturbations ($[g_j]=0$) when $n \to \infty$, and RG analysis can be controlled by the parameter $\epsilon=2/n$. By integrating out the fast modes belonging to the shell $E_c e^{-l}<\sqrt{\omega^2+v^2 k^2_z}<E_c$ and $0<k_\perp<\infty$, and subsequent rescaling according to $\omega \to e^{-l} \omega$, $k_z \to e^{-l} k_z$, $k_\perp \to e^{-2l/n} k_\perp$ and $\Psi \to e^{(3+\frac{2}{n})l/2} \Psi$, we obtain the following RG flow equations
\allowdisplaybreaks[4]
\begin{align}
\beta_{\lambda_0}&= -\epsilon \lambda_0 + {\mathcal O} \left( \lambda_i \lambda_j/n\right), \nonumber \\
\beta_{\lambda_2}&= -\epsilon \lambda_2 + \lambda^2_+-\lambda^2_3+ \lambda_0 (\lambda_3-\lambda_+) + {\mathcal O} \left( \lambda_i \lambda_j/n \right), \nonumber \\
\beta_{\lambda_3}&= -\epsilon \lambda_3 + (2 \lambda_3-\lambda_0) (\lambda_3-\lambda_+)+ {\mathcal O} \left( \lambda_i \lambda_j/n \right), \; \nonumber \\
\beta_{\lambda_4}&= -\epsilon \lambda_4 + (\lambda_+ -\lambda_3)^2 + {\mathcal O} \left( \lambda_i \lambda_j/n \right), \label{RG_epsilon}
\end{align}
to the leading order in $\epsilon$, where $\lambda_j= g_j \left(E_c\right)^{\epsilon}/(4 \pi^3 \alpha^{\epsilon}_n)$ for $j=0,2,3,4$ are dimensionless coupling constants, $\lambda_+=\lambda_2+\lambda_4$, $\beta_x=\frac{dx}{dl}$ and $E_c$ is the ultraviolet energy cut-off for Weyl fermions. The contributions $\sim 1/n$ capture the \emph{sub-leading} divergence, which of course vanish for the special limit $n \to \infty$~\cite{supplementary}.

The RG flow equations support following fixed points in the four dimensional space of $(\lambda_0, \lambda_2, \lambda_3, \lambda_4)$. The attractive fixed point located at $(0,0,0,0)$ represents a stable non-interacting WSM phase. For $n \to \infty$ the two other fixed points (repulsive or QCPs) are located at $(a)=(0,-1,2,1)\frac{\epsilon}{4}$, describing a continuous QPT into the AN phase, and $(b)=(0,1,0,1)\frac{\epsilon}{4}$, which, on the other hand, possesses an $O(4)$ symmetry (since $\lambda_2=\lambda_4$ and $\lambda_0=\lambda_3=0$ at this QCP). When $n \to \infty$, the in-plane kinetic term $\propto k^n_\perp$ drops out of $H^n_W$. The resulting Hamiltonian (effectively one-dimensional) anticommutes with $\Gamma_{01}$, $\Gamma_{02}$, $\gamma_0$ and $\Gamma_{05}$, giving rise to a \emph{spurious} $O(4)$ symmetry, which is actually absent for any WSM with finite $n$. After accounting for the subleading $1/n$ corrections~\cite{supplementary}, the flow equations still support only two QCPs and the new location of QCP $(b)$ is given by
\begin{align}
(b) &\approx \bigg( 0,\frac{1}{4}-\frac{0.33}{n}\delta_{n,2p}-\frac{0.28}{n}\delta_{n,2p+1}, \frac{0.94}{n}\delta_{n,2p+1}, \nonumber \\
& \frac{1}{4} -\frac{0.17}{n}\delta_{n, 2p} +\frac{0.12}{n} \delta_{n, 2p+1} \bigg)  \epsilon,
\end{align}
for large $n$, where $p$ is an integer. Upon investigating the flow of various order parameter susceptibilities, we find this QCP describes a continuous QPT to the AI phase. We note that the $1/n$ corrections do not alter the nature of QCP $(a)$, apart from causing a non-universal shift in its location~\cite{supplementary, susceptibility}. A fully unstable fixed point separates the domain of attraction of these two QCPs [magenta point in Fig.~\ref{competing_PD}(upper panel)]. A schematic RG flow diagram and the resulting cuts of the phase diagram in various coupling constant planes are shown in the upper and lower panels of Fig.~\ref{competing_PD}, respectively.

Next we discuss the scaling phenomena at these two QCPs. From the leading order $\epsilon$-expansion we find the CLE at each QCP to be $\nu^{-1}=\epsilon$. The fact that the CLE is equal at all QCPs is a consequence of leading order perturbative analysis. CLEs at different QCPs are expected to become different once the higher order corrections are taken into account. Notice that we recover the mean-field exponent $\nu=\frac{1}{2}$ for $n=1$ WSM~\cite{zinn-justin}. By contrast, for $n=2$ and $3$, we respectively obtain $\nu=1$ and $3/2$, suggesting that general WSMs with $n>1$ can support rare examples of \emph{non-Gaussian} QCPs in three dimensions. As a direct consequence of non-Gaussian criticality, for example the spectral gap in the AI phase ($\Delta_{AN}$) does not possess any \emph{logarithmic correction} for any $n>1$~\cite{supplementary}.

\emph{First-order transition}: Finally, we address the effects of generic short range interactions in the vicinity of the WSM-BI QCP, located at $\Delta=0$. The analysis is carried out for two-component critical excitations ($\psi$) for which local interactions are captured by four terms $H^{1}_{int}= \tilde{g}_\mu \int d^3x \left( \psi^{\dagger} \sigma_\mu \psi \right)^2$, where $\mu=0,1,2,3$. Due to the \emph{Fierz identity} of Pauli matrices~\cite{supplementary}, only one of the four coupling constants is independent, which we choose to be $\tilde{g}_3$. At the non-interacting QCP, the spectra of critical excitations are $E({\mathbf k})=\pm \sqrt{\alpha^2_n k^{2n}_{\perp}+B^2 k^4_z}$, giving rise to strong anisotropy between in-plane and out-of-plane response functions for any $n$ and the low-energy DOS $\varrho(E) \sim |E|^{\frac{4-n}{2n}}$. The scaling dimension of the coupling constant $[\tilde{g}_3]=-\frac{4-n}{2n}$ suggests that weak short-range interaction is an \emph{irrelevant} perturbation at the WSM-BI QCP for any $n<4$, thus only renormalizing the phase boundary between the WSM and BI [see Fig.~\ref{firstorder}(b)].

The effect of electronic interaction at the WSM-BI QCP can quantitatively be demonstrated from the following mean-field free-energy density~\cite{supplementary}
\begin{align}~\label{freeenergy}
F=\frac{\Sigma^2}{2 \tilde{g}_3}- \int \frac{d^3{\mathbf k}}{(2 \pi)^3} \sqrt{\alpha^2_n k^{2n}_\perp + \left( B k^2_z + \Delta + \Sigma \right)^2 },
\end{align}
obtained after Hubbard-Stratonovich decoupling the four-fermion interaction in favor of a bosonic field $\Sigma=\langle \Psi^\dagger \sigma_3 \Psi \rangle$ and subsequently integrating out critical fermions. We numerically minimize the free-energy to arrive at the phase diagram, shown in Fig.~\ref{firstorder}(b), for $n=1$. The salient features of the phase diagram can be appreciated by expanding the dimensionless free energy in powers of $\rho$, yielding
\begin{equation}~\label{free-expand}
f \approx \left[\frac{m^2}{2 \lambda} + f_0(n)\right] + \rho \left[ \frac{m}{\lambda} + f_1(n)\right] + \sum_{j \in p} \rho^j f_j (n),
\end{equation}
after shifting $\rho+m \to \rho$, which contains all odd powers of $\rho$, where $\lambda=\tilde{g}_3 E^{\frac{4-n}{2n}}_\Lambda/a_n$, $\rho=\Sigma/E_\Lambda$, $f=F a_n E^{-\frac{4+3n}{2n}}_\Lambda$ and $a_n=[2 n \alpha^{2/n}_n B^{1/2} (2 \pi)^2]$. For weak enough interaction (no condensation of $\rho$), the profile of $f$ contains a single global minimum and the second term in Eq.~(\ref{free-expand}) defines the phase boundary $m=-\lambda f_1(n)$ between the WSM and BI, without altering the nature of the transition. However, beyond a critical strength of interaction ($\lambda>\lambda_\ast \sim \frac{4-n}{2n}$), the $\rho$ field acquires an expectation value and all odd powers of $\rho$ in the free-energy become important, and $f$ contains two inequivalent local minima. Consequently the direct transition between the WSM and BI at strong coupling becomes a \emph{fluctuation driven first order transition}~\cite{roy-goswmai-sau}. As shown in Fig.~\ref{firstorder}(a), the direct transition between the WSM and BI can be eliminated by an intervening BSP (the gray region). The BI-BSP transition is continuous.

\emph{Conclusions}: To summarize, we have shown that a strongly interacting general WSM can either undergo continuous QPTs into a translational symmetry breaking AI or a rotational symmetry breaking nematic (namely AN) phase or be separated from a symmetry preserving BI by a first order transition, as shown in Figs.~\ref{firstorder} and \ref{competing_PD}. Using a perturbative RG analysis controlled via an $\epsilon=2/n$-expansion, with $n$ being the monopole charge of the Weyl point, we establish that the CLE at the WSM-BSP QCPs is $\nu=1/\epsilon$, suggesting non-Gaussian quantum criticality for any $n>1$~\cite{footnote-2}. All BSPs can display true long range order and undergo a genuine thermal phase transition to WSMs. While the specific heat in a general WSM scales as $C_v \sim T^{\frac{2}{n}+1}$, inside the nematic phase $C_v \sim T^3$. Consequently the Gr$\ddot{\mbox{u}}$neisen ratio ($\Gamma_G$) in the WSM and nematic phase respectively diverges as $\Gamma_G \sim T^{-\left(\frac{2}{n}+2 \right)}$ and $T^{-4}$. By contrast, at the WSM-BI QCP $C_v \sim T^{\frac{4+n}{2n}}$ and $\Gamma_G \sim T^{-\frac{4+3n}{2n}}$. For small frequencies ($\Omega$) the dynamic conductivity ($\sigma$) scales as $\sigma_{zz}\sim \Omega^{\frac{2}{n}-1}$ and $\sigma_{jj} \sim \Omega$ for $j=x,y$, with $\sigma_{xx}=\sigma_{yy}$ (due to rotational symmetry in the $xy$ plane) inside the WSM phase. But, in a nematic phase $\sigma_{jj} \sim \Omega$ for $j=x,y,z$, with $\sigma_{xx} \neq \sigma_{yy}$ due to the lack of rotational symmetry. In contrast, the conductivity for the AI phase shows \emph{activated} behavior. Therefore, the onset of either AI or the gapless nematic phase can be identified through various thermodynamic and transport measurements.

The application of external strain can further enrich the phase diagram of interacting Weyl materials. For example, when the external strain is weak it couples with the nematic order as an external field and splits the Weyl node of strength $n$ into $n$ copies of simple Weyl points with $n=1$. Subsequently, such a nematic WSM can undergo a continuous QPT into an AI phase for strong backscattering, as suggested by the phase diagram in Fig.~\ref{competing_PD}. A detailed analysis of the interplay of strain and electronic correlations in Weyl materials is left for future investigation. 

\emph{Acknowledgement:} This work was supported by NSF-JQI-PFC and LPS-MPO-CMTC (B. R. and P. G.), and partially by Welch Foundation Grant No.~C-1809, NSF CAREER Grant no.~DMR-1552327 of Matthew S. Foster (B. R.). B. R. is thankful to Nordita, Center for Quantum Materials, for hospitality.


\begin{thebibliography}{}
\bibitem{spectrum} Quasiparticle spectrum inside a BSP is obtained by diagonalizing effective single particle Hamiltonian $H^n_W+H_j$, where $j=AI, RN, AN$.
\bibitem{footnote-1} $\left\{H_{AI}, \gamma_5 \right\}=0=[H_{AI}, \Gamma_{12}]$ dictates that AI breaks (preserves) translational (rotational) symmetry, while $\left\{H_j, \Gamma_{12} \right\}=0=[H_j, \gamma_5]$ imply that nematic orders ($j=AN, RN$) break (preserve) rotational (translational) symmetry.
\bibitem{footnote-2} Onset of BSPs can also take place through the instability of strongly interacting critical modes residing at the WSM-BI QCP ($\Delta=0$) through continuous QPTs across the multicritical point (red dot), shown in Fig.~\ref{firstorder}(a). The leading order CLE for such QPTs is $\nu=\frac{2n}{4-n}$, indicating its non-Gaussian nature for any $n<4$.
\end{thebibliography}
\end{document}